# Twin nucleation in Ti: A study using nudged elastic band (NEB) method

Deepesh Giri[a,b], Haitham El Kadiri[a,b,c], Christopher D. Barrett[a,b]

[a]Center for Advanced Vehicular Systems, MS 39759, USA
[b]Department of Mechanical Engineering, Mississippi State University, MS 39762, USA
[c]Universite Internationale de Rabat, Rabat-Shore Rocade Rabat-Sale, Morocco

**Abstract:**
Capturing twin nucleation in full-field crystal plasticity is a long-standing problem in materials science. The challenge resides mainly in the biased regional lattice transformation associated with twin formation in defiance of its obedience to a threshold stress law which could be fulfilled in regions where twinning is deferred. Hence, determining a favorable site for nucleation of a twin variant remains a daunting task. We hypothesized that this site-specific nucleation is sensitive to the prior atomic structure of the lattice so twin embryos form in regions where the lattice transformation energy is minimum. Thus, quantifying the local strain energy required to trigger a stable twin underscores the non-pseudo-slip behavior of twin nucleation and growth. We performed atomistic calculations based on the nudged elastic band method to identify the minimum energy path and activation energy associated with {10-12} twin nucleation in titanium. Results of calculations demonstrate that the role of stress and atomic structure in twin nucleation could be understood in terms of the minimum energy path, energy barrier, and relaxed energy. Remarkably, for symmetric tilt grain boundaries, a linear correlation between the nucleation stress and grain boundary energy was observed.

**Introduction**
Twinning is a prominent plastic deformation mechanism for many commercial materials such as TWIP steels and other hexagonal close packed (HCP) metals (Mg, Ti, etc.) [1, 2, 3]. These HCP metals are widely used in industrial applications because of their low density and high stiffness. However, because they lack sufficient independent slip systems, twinning plays significant role in their strength and ductility. Twinning, along with pyramidal slip provides a means to accommodate deformation along c-axis in HCP materials. Contrary to face centered cubic (FCC) and body centered cubic (BCC) metals which have limited number of twinning systems, there are at least seven twinning modes in HCP metals, most of which are observed to be active in Ti and Zr [1, 4, 5]. Of particular interest is {10-12} twin mode which is highly prevalent in HCP metals [3].

Twinning is a much more complex form of plasticity than slip. Twinning alters the positions of atoms, induces shear strain in a particular region inside a material, shuffles [6] the atoms and generates new interfaces. Twinning should be viewed as a combination of 3 steps – nucleation [7, 8], growth [9] and propagation [10]. Nucleation and growth of twins is highly dependent on the points of their origin. Propagation of twins is a complex phenomenon as it involves twin-twin interactions [10], faceting [11] and patterning. This occurs very quickly as twins can shoot from one side of a grain to the other and in Mg such twins have been known to grow until the parent grain is entirely consumed by the twin [12]. For adequate combinations of stress sign and state, {10-12} twinning could take on a large part of plastic accommodation in traditional hexagonal close packed (HCP) and other low symmetry metals [1, 3], leading to a strong anisotropy and asymmetry known to be associated with poor ductility and low energy absorption of the material. Such an unusual process requires dedicated treatment in higher scale modelling and often can only roughly approximate experimental data as the material description changes from that which is used at lower scales. This work, within the frame of Molecular Statics (MS), focuses on the origin of {10-12} twins in Ti.

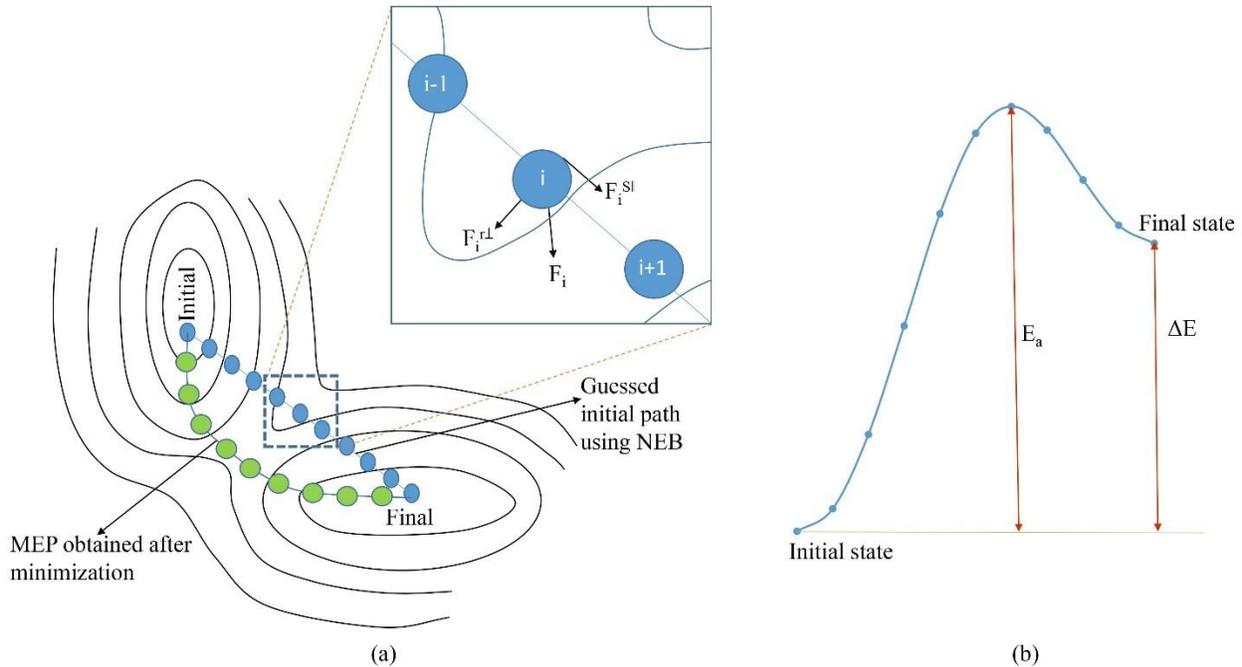

**Figure 1:** (a) A schematic representation of nudged elastic band (NEB) method. The initial and final stable configurations are shown to lie on a potential energy surface (PES). The NEB method at first does a linear interpolation between the two stable states (blue dots) and then minimizes the intermediate states to provide the minimum energy path (green dots) [13]. (b) A typical graph depicting the minimum energy path (MEP). The dots represent energy of each of the intermediate states. $E_a$ is the height of activation barrier while $\Delta E$ is the energy difference between the two stable states of the transition.

Twin nucleation carries a great significance since the subsequent growth of twins and eventually the twinned microstructure is highly dependent on the points of their origin. Twin nucleation mechanisms for HCP metals are broadly divided into homogeneous and heterogeneous nucleation models. Homogeneous model, which requires pristine parent domain, was first introduced by Orowan [14] where he derived an expression for minimum stable size of twin nucleus in Zn. Heterogenous nucleation occurs at defect sites such as grain boundaries (GBs) or dislocations [15, 16, 17, 18, 19]. Identification of the origin points for twin nucleation in an initial microstructure is still at early stages of research, making it a ripe area for dedicated study.

Determining where the first twin nucleus forms is an unresolved issue although Paudel et al [20], Barnett et al [21] and Siska et al [22] have done some work to estimate the nucleation of next twin based on the stress state of the first one. We tried to quantify twin formation susceptibility based on energy criteria in our previous works [23, 24]. Owing to the importance of origin points for twin nucleation, a guideline is required on how to address the tendency of certain sites to nucleate twins. The effect of misorientation [25], angle of the interface, and the role played by vacancies and segregated elements in the GB need to be well understood. The process of twin nucleation has a strong non-Schmid behavior [26]. Hence the effects of applied stress tensor and the response of various sites to different types of loading need clear understanding before we can rate the propensity of sites to twin nucleation. Atomistic simulations conducted by J. Wang [27, 28, 8] Hong-lu [29], Barrett [30] have attempted to address these issues with some success but the fact that MD is conducted at such a small size and time-scale and hence can't sample diffusion effects stands as a major hindrance. To answer these questions and quantify the points where twins prefer to nucleate, we turn to atomistic simulations with Nudged Elastic Band (NEB) method [31, 32].

**NEB Method**
NEB method is an efficient technique to find minimum energy path (MEP) between two stable states of a transition [31, 32, 33]. MEP is a curve through phase space that finds its way over saddle points to connect intermediate states using the lowest energies possible to create that curve [34]. A MEP is a continuous line describing the evolution of reaction and the intermediate states between reactants and products. MEP emerges from reactants, traces the intermediate stages, reaches the saddle before terminating in products. Any point on the MEP is at an energy minimum in all directions perpendicular to the path [13]. The highest point on the MEP is known as the saddle point energy and gives an estimation of the activation barrier and reaction rate for that particular reaction. A typical energy plot with 11 images depicting MEP is shown in Figure 1(b).

NEB is a chain of states method [35] that uses a string of images (duplicates or replicas) to describe transition from an initial configuration to final configuration. It is an algorithm which is an approximation defined to find the MEP. In other words, NEB describes the movement of atoms from initial stable state to final stable state while requiring the least amount of energy possible. There could be several reaction pathways connecting the reactants and products out of which NEB identifies the one that requires minimum energy. The NEB algorithm guesses an initial path for all the atoms between the two local minima on the potential energy surface (PES). Then it iterates several times to relax this initial path for optimizing the location of the images and finally gives the MEP along with the height of activation barrier. Figure 1(a) is a schematic of NEB method on a PES. A PES is a mathematical function that gives energy of a system as a function of its geometry.

To govern the identification of MEP, NEB method employs a set of images/replicas (typically, 4-20) of the system between initial and final configurations [36]. The number of images to be used in NEB seems to be a tricky choice between getting a high resolution of the energy landscape and the computational requirements. Sobie et al [37] used 30-60 images in their calculations whereas Henkelman et al [32] opines that a band with more than 12 images takes much longer to converge and develops large fluctuations before eventually converging to the MEP. The NEB method ensures continuity of the MEP by providing spring interactions between the images along the band. The spring forces constrain the images by maintaining their distances from neighboring states from growing too large. This enables all the states to relax while remaining continuously connected so that it can construct a path through configuration space.

Each image i experiences forces due to spring interactions (spring forces) and the gradient of potential energy (true force). This force projection is referred to as 'nudging'. Mathematically,

$$\vec{F_i} = \vec{F_i^s} + \vec{F_i^r} \qquad 1)$$

Where $F_i^s$ is the spring force and $F_i^r$ is the true force. The perpendicular component of the spring force causes drift of the energy pathway while the parallel component of true force allows non-equidistance distribution of images along the energy pathway. In the NEB method, these two forces are projected out while minimizing the elastic band. Hence the total force acting on an image i is the sum of the spring force along the band and the true force perpendicular to the band. Now, the force on image i becomes:

$$\vec{F_i} = \vec{F_i^{s\parallel}} + \vec{F_i^{r\perp}} \qquad 2)$$
$$= k[(\vec{R}_{i+1} - \vec{R_i}) - (\vec{R_i} - \vec{R}_{i-1})].\hat{\tau}_{\parallel}\hat{\tau}_{\parallel} - \vec{\nabla}V(\vec{R_i})_{\perp}$$
$$= \vec{F_i^s}.\hat{\tau}_{\parallel}\hat{\tau}_{\parallel} - [\vec{\nabla}V(\vec{R_i}) - \vec{\nabla}V(\vec{R_i}).\hat{\tau}_i]$$

where $\hat{\tau}_{\parallel}$ is a unit vector tangent to the minimization path, R is position of image i and k is the spring constant. The choice of the spring constant has negligible effect on computational time and no effect on the activation energy values [37].

The NEB method is divided into two stages, both of which are performed using damped dynamics minimization schemes. In the first stage, the replicas converge towards the MEP while in the second stage, barrier climbing calculations are done to drive replica with the highest energy to the top or saddle point of the barrier. After a few iterations with the initial NEB, the climbing image calculations provide a rigorous convergence to saddle point without any significant additional computation costs [36].

The NEB method has been successfully applied to a wide range of problems - surface diffusion of atoms, dislocation nucleation, chemical reaction in enzymes etc. [38, 39, 40, 41]. There was 'Plain Elastic Band' (PEB) method, that preceded NEB. PEB faced problems of corner cutting with higher stiffness and lower resolution at the saddle point with lower stiffness. NEB overcomes these shortcomings by using force projection scheme to converge the elastic band to the MEP [31].

There are other available schemes which converge the elastic band to the MEP. Climbing image NEB (CI-NEB) method is very similar to NEB which identifies image with highest energy and moves it up the potential energy surface. It provides rigorous convergence to saddle point and has different spacing of the images on each side of the climbing image. Variable spring constants can be employed in the NEB or CI-NEB schemes to obtain better resolution near the saddle point [36]. Doubly nudged elastic band method (D-NEB), string method and simplified string method [13] are all similar to the NEB method except for a few modifications. An automated NEB (AutoNEB) method is another algorithm that is like regular NEB but uses fewer computational resources [42].

**Methodology**

This paper focuses on NEB calculations on {10-12} twin formation in Ti. Some calculations were also performed with Mg for comparison. We first generated a bicrystal in both Mg and Ti to compare the twinning process. The bicrystal structure and misorientation varies slightly from Ti to Mg because of the difference in their lattice parameter ratios, but Mg and Ti exhibited very similar results for the {10-12} twin mode [43, 44]. A detailed discussion of the results for the Ti bicrystal are published in one of our earlier works [23]. For that and new results herein, a meam/spline Ti potential developed by Henning et. al [45] was used with LAMMPS [46]. The obtained data was further minimized to get initial stable configuration.

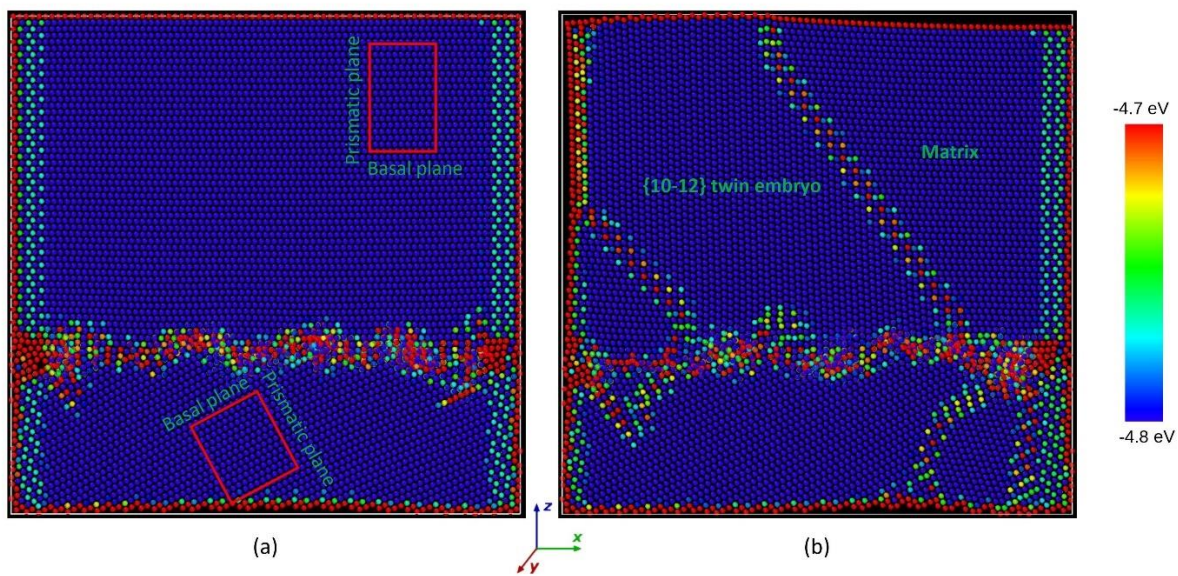

**Figure 2:** [-2110] axis asymmetric tilt GB. The bicrystal contains 48876 atoms and 740 vacant sites. (a) Initial stable configuration (b) Final configuration, stabilized under 500 MPa compression, with {10-12} twin embryo. Color coding is done based on energy in Ovito. A color bar on the right depicts energy spectrum between -4.8 eV and -4.7 eV. Although only the upper twin was inserted, another twin also spawned on the lower right corner during minimization. The GB looks distorted due to the presence of significant number of vacancies.

A Mg bicrystal containing 188600 atoms and [-2110] axis asymmetric tilt GB with around 30° misorientation (minimum angle required to rotate a grain to another grain) was created in

Matlab. It was generated by appending two low-index planes to make a low-misorientation tilt bicrystal. This type of MD simulation has been demonstrated in the past to give good nucleation results, and experimental work has confirmed that twin nucleation is more frequent at boundaries with these misorientations [25]. Moreover, the boundary energy is high, as most HCP tilt boundaries with similar misorientations are, so the boundary adopts a somewhat complex core structure upon relaxation [47, 48]. We inserted a {10-12} twin embryo onto this bicrystal using Matlab code which solves for the displacement field inside the embryo using isotropic approximation [49]. The positions of the mapped atoms are computed such that the atom identifiers all match the initial atom positions correctly. The twin embryo is enclosed by disconnection loops generated as dislocation segments. The total displacement inside the embryo is a result of the combined effect of all the dislocation loops and shuffle [6], while atoms outside are displaced only due to the dislocation loops. This boundary is chosen for NEB analysis partially because it does produce twin nucleation with dynamics, which matches quite nicely with the structure produced by NEB. NEB is run on this bicrystal using modified embedded atomic method (MEAM) potential developed by Wu et. al [50].

The twin embryo is inserted on the upper half of the initial state to get final state, which is further minimized to get final stable configuration as shown in Figure 2. During minimization, certain amount of stress is applied to hold the twin embryo stable. The value of this stress was determined purely on a trial-and-error basis. We generally started energy minimization with 2 GPa applied stress and kept on reducing it until the twin is no longer stable. The lowest value of stress is then chosen for work done and energy relaxations. During relaxation, except the applied stress, all other stress components were reduced as close to zero as possible.

Since stability of a twin embryo is influenced by its size, we insert a twin embryo that spans more than (1/3)$^{rd}$ volume of the bicrystal to ensure it doesn't collapse immediately. These twins are easily produced by tension along or compression perpendicular to the c-axis during plastic deformation [1]. As the upper half of the bicrystal has c-axis along z direction, we apply horizontal compressive stress. We fix a few layers of atoms on each side of the simulation box in x dimension to mimic real stress application scenario. Using these stable configurations, we run NEB calculations to obtain MEP for this transition. The boundary conditions used are surface, periodic and surface in the x, y and z dimensions respectively. Conjugate Gradient (CG) minimization algorithm is used for minimizing these files while dampened dynamics is used for the NEB calculations. We run initial NEB for 12000 timesteps while barrier climbing NEB is run for another 9000 timesteps with both force and energy tolerance of $10^{-16}$ units. The number of timesteps were compared with higher values up to 100,000, but we found that the MEP was barely altered by these higher values.

Convergence of the barrier climbing with more timesteps was not needed because we measured the distance through phase space from one replica to another and found them to be very close to evenly spaced. We use a very short timestep of 0.1 fs which means our NEB simulations run for a total time of 2.1 ps. Spring constants of 0.1 eV/Ang$^2$ are used for both parallel and perpendicular nudging forces. The obtained data is visualized and analyzed using Ovito [51].

After a number of convergence studies, we selected 24 replicas to perform the majority of NEB calculations. This is slightly more than the 5-20 used by Henkelman and Jonsson [32]. We used [-2110] axis asymmetric tilt GB in Figure 2 and varied the number of replicas from 8 to 150 to check if this variation has any effect on the MEP. In all the cases, height of the energy barrier and shape of MEP were almost the same indicating minimal effect of the number of replicas.

Following the proof-of-concept work done with the initial bicrystal which compared the effects of material, replicas, timesteps, etc., we studied the role of grain boundary misorientation on ease of twin nucleation. We focused on the role of symmetric tilt grain boundary (STGB) and few other frequently occurring GBs on {10-12} twin nucleation in Ti. GBs can be classified into various groups based on the details of their structures. Dislocations, burgers vector, step height, tilt angle, misorientation etc. are the characteristics of GBs depending on which their ability to absorb, annihilate or release the defects are defined. GB misorientations play significant role on the nucleation, growth and propagation of {10-12} twins. Hence, it is crucial to understand the GB misorientation distribution [52, 53, 54]. This work discusses STGBs and their influence on twin nucleation phenomenon.

We employed [-2110] axis STGBs for this work. STGBs are among the simplest of all GBs and can be described only with the tilt axis and tilt angle. They have relatively low energies and hence are more stable. However, for a given STGB, depending on the tilt angle, the GB structure can vary resulting in the variation of GB energies. We generated a number of [-2110] axis STGBs by varying the tilt angle from 0 to 90 degrees. Figure 3 is one such STGB with 61.56° tilt angle.

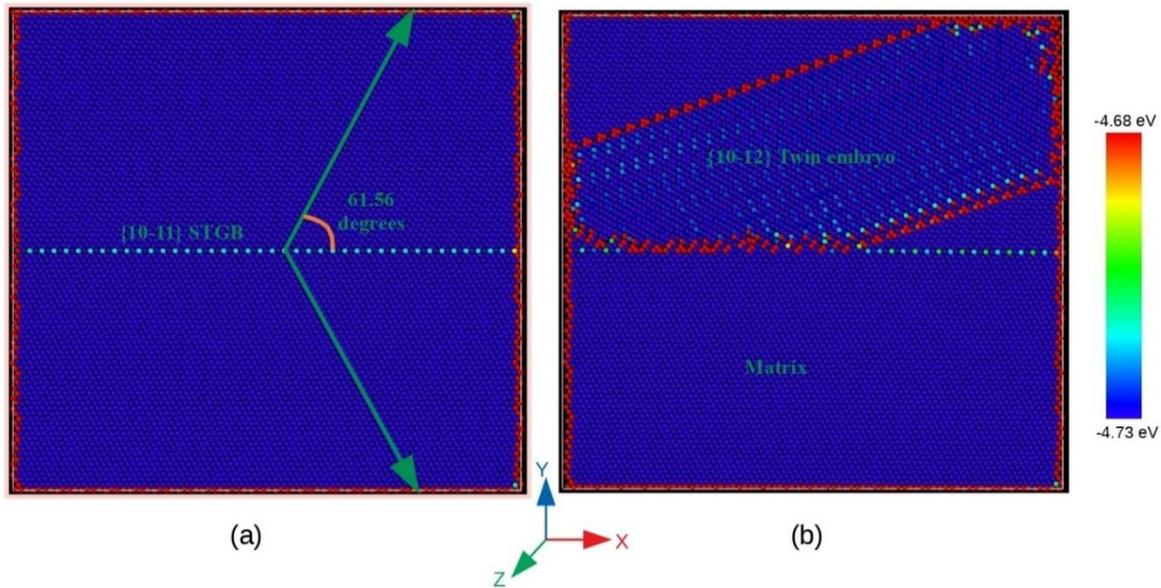

**Figure 3:** [-2110] axis symmetric tilt grain boundary (STGB) with 61.56° tilt angle, 47880 atoms and surface, surface, periodic boundary conditions. (b) {10-12} twin embryo inserted on the upper half of (a). Color coding is done based on energy in Ovito. A color bar on the right depicts energy spectrum between -4.73 eV and -4.68 eV. Color varies from red for higher energy atoms to blue for lower energy atoms.

These bicrystals are then minimized to obtain relaxed initial structures. A {10-12} twin embryo is inserted on their upper half and minimized to obtain final stable structures. Energies of these GBs are calculated and tallied against the values obtained by Wang et. al [47]. Individual GB energies and the overall plot closely match with this work as can be seen in Figure 6. Then we run NEB calculations to quantify the effects of such STGBs on {1012} twin nucleation.

We computed energy released during twin nucleation in each case and obtained the values of minimum stress required to stabilize the twin embryo for positive energy relaxation. This provided a very simple relationship between applied stress and GB energy. $E_{work}$ is the amount of energy added to the simulation by the imposed loading during twinning, provided stress is kept constant. It is calculated as:

$$E_{work} = Force * displacement \quad\quad\quad 3)$$
$$= (Stress * Area) * displacement$$

where, Stress is the applied stress, Area is the area of the plane where stress is applied and displacement is the average contraction of the simulation box that occurred during twinning. The energy released by twinning is then estimated as:

$$E_{twin} = E_{work} - \Delta E \quad\quad\quad 4)$$

where ΔE is the change in stored energy from the initial to final state. We also measured twin thickness and identified its relationship with stabilizing stress.

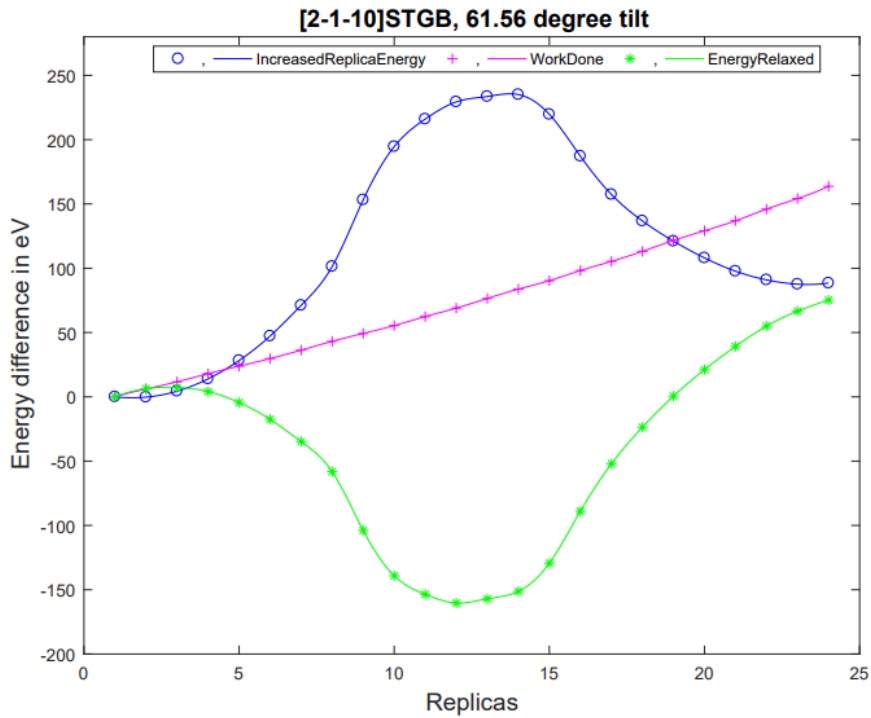

**Figure 4:** Minimum energy path (MEP) during twin formation on a Ti bicrystal with [-2110] axis symmetric tilt grain boundary (STGB) and 61.56° tilt angle. The process has an energy barrier of around 230 eV and energy relaxation of around 70 eV.

## Results

We ran NEB simulations on a number of [-2110] axis STGBs and obtained the various energy parameters. Here we present comparative plots of MEP, work done and energy relaxed during twin formation in 61.56°, and 70.11° tilt angle boundaries, the results of which are in Figures 4, and 5 respectively. We computed these values for a dozen other GBs but we couldn't obtain distinct energy barrier for all. Some of the processes returned negative values of $E_{twin}$. This indicates configuration in which twinning would not occur as twin nucleation is an exothermic process. Hence in such cases, we re-computed the minimum amount of stress required to get a positive value of $E_{twin}$. Since the stress required to twin is more useful than the energy barrier, we tried to tailor the stresses such that the energy released by twinning was near zero, indicative of the critical resolved shear stress (CRSS).

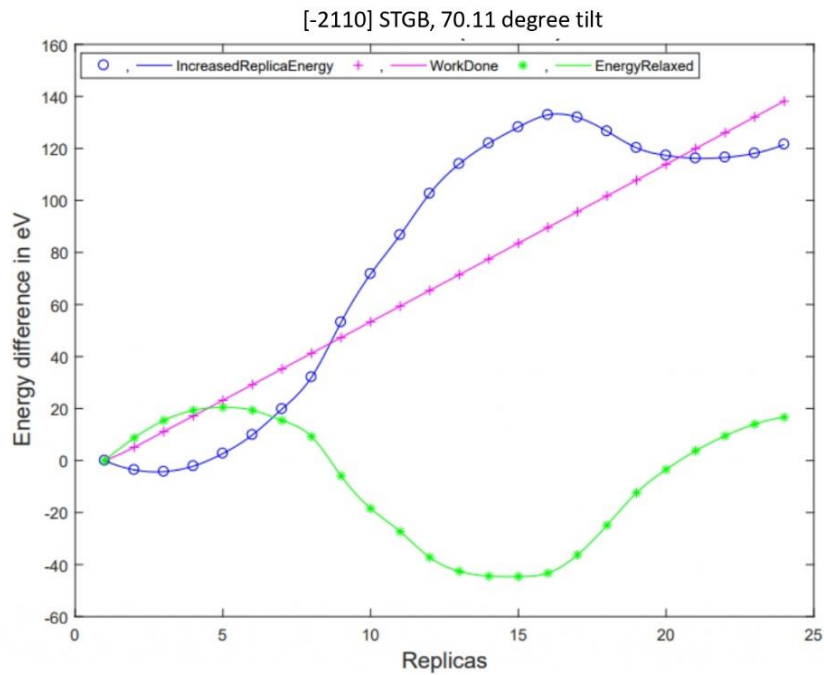

**Figure 5:** Minimum energy path (MEP) during twin formation on a Ti bicrystal with [-2110] axis symmetric tilt grain boundary (STGB) and 70.11° tilt angle. The process has an energy barrier of 130 eV and energy relaxed is around 20 eV.

We plotted the values of stabilizing stress against the corresponding GB energies, the result of which are in Figure 6. The Y-axis in the graph resembles GB energy in $mJm^{-2}$ and stress in MPa. The graph indicates that both values are showing similar patterns with respect to tilt angles. Roughly, we have higher values of stress for higher GB energies and vice-versa. Another interesting observation was that all the energy cusps have lower stress requirements than other GBs. Upon plotting these stress values against misorientation angles, another meaningful pattern was observed. As seen in Figure 7, for misorientation angles in the range of 0-30° and 60-90°, twin nucleation becomes more difficult with increase in misorientation angle due to increase in stress. However, in the 30-60° range, much lower values of stress are required. The GBs with energy cusps lie within this range.

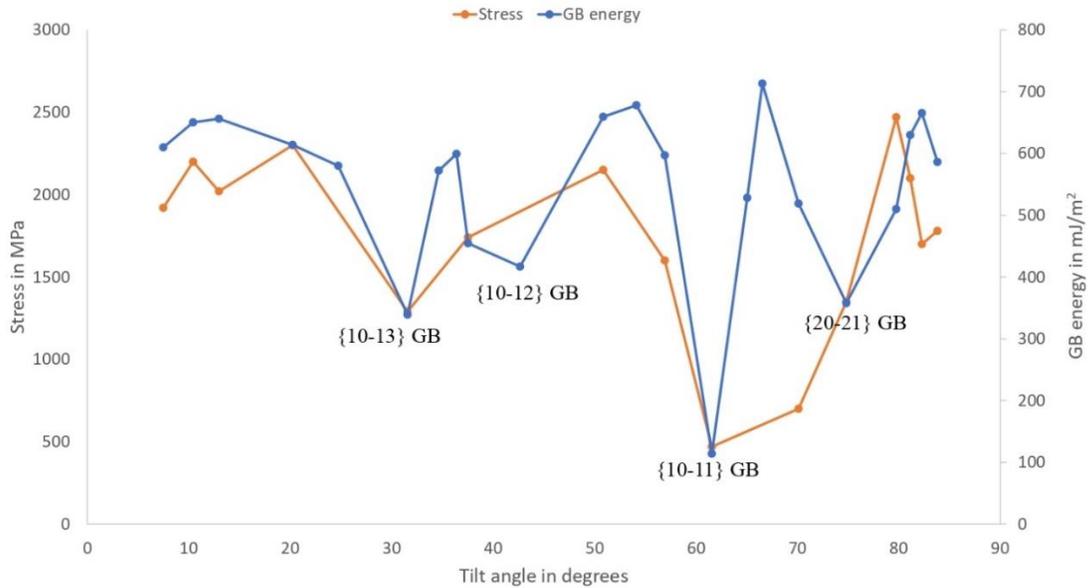

**Figure 6:** The required amount of stabilizing stress seems to follow the same path of grain boundary (GB) energy with respect to tilt angle. The cusps represent lower values of GB energy and stabilizing stress compared to others.

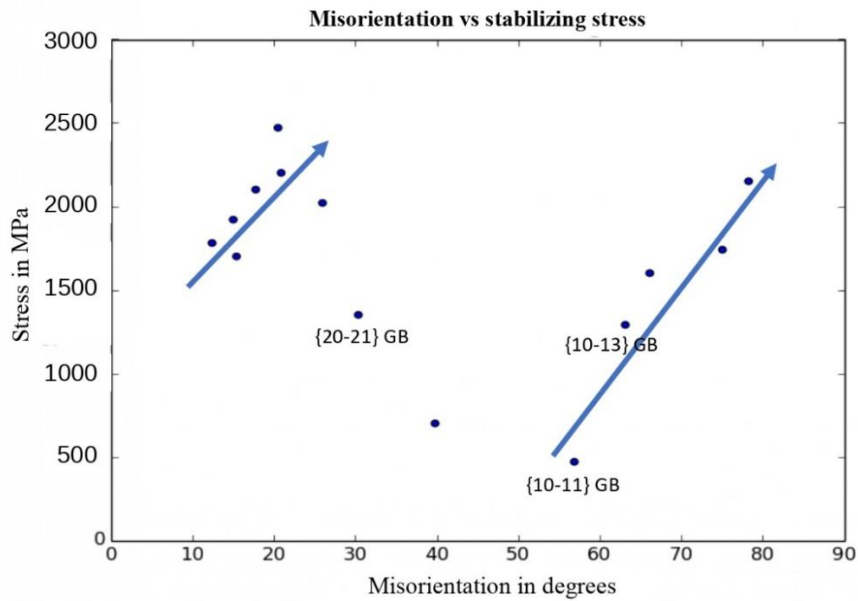

**Figure 7:** Stabilizing stress increases with misorientation angle except in 30-60° range, making it the most suitable for twin nucleation. Misorientation angle is the minimum amount of rotation required to orient one grain to another.

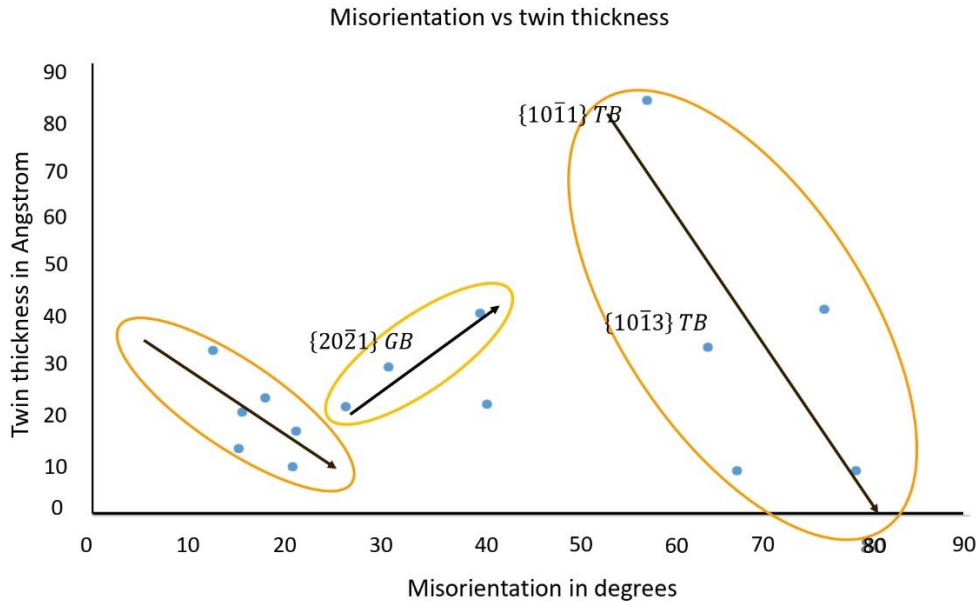

**Figure 8:** Twin thickness decreases with misorientation angle except in the 30-60° range. This is consistent with results in Figure 7 and indicates that it is a favorable region for twin growth.

We also investigated the thickness of our twins and plotted them against the misorientation angles. Figure 8 depicts the relationship between these two parameters. Twin thickness shows a decreasing trend for misorientation angles in the range of 0-30° and 60-90° and increasing trend in 30-60° range. The behavior shown by stress and twin thickness are consistent because higher stress is required where twin nucleation is difficult which means twins are more likely to be thinner, and vice-versa.

**Discussion**
Some of the MEPs (e.g. Figure 5) have few replicas on the right of the replica with minimum energy. The reason could probably be that despite our energy relaxations, the final configuration was not fully relaxed and hence NEB found some intermediate configurations with lower energy. It could also be a resolution issue. NEB locates local minimum based on the number of available replicas and in this case maybe 24 was not enough to find out deeper minimum. Employing larger number of replicas in such case may help to locate global minimum. And for cases where we don't get a distinct energy barrier, increasing the dimensionality in phase space might help to locate the saddle points and identify the barrier. The MEP is a path through phase space which is spanned by all the degrees of freedom of all the coordinates of the atoms. It is a very high dimensional space. By increasing the number of replicas, we can expect to get more local minimums and maximums.

NEB script uses both energy/force and timestep cutoff to decide on the number of calculations. In our case, the timestep cutoff was reached first, which means forces greater than $10^{-16}$ still existed in the system. This does not merit concern because we have also done convergence studies with much greater timestep cutoffs but the effect of that on the MEP was negligible.

Table 1 presents magnitudes of slip vectors [55] during various stages of the twin nucleation process. 4 atoms were selected from each of the twinned, untwinned and GB regions of the bicrystal and their slip vectors considering 12 nearest neighboring atoms were calculated. It is clear from the table that the atomic movement in twinned region is below 1 Angstrom while in the untwinned region it's not even 0.1 Angstrom. Maximum movement is observed in the GB region but that hardly exceeds the burgers vector, thus backing our approximation that the atoms haven't moved so much during the twinning process. This suggests that the environment is not drastically changing from initial to final configurations and we can be fairly confident that our linear starting approximation works well to generate the MEP. The atomic movements in the order of fraction (about 1/6 - 1/3) of the lattice parameter clearly shows that the path through phase space is simple and hence fewer replicas (24) and a linear interpolation both work well to study this phenomenon. On the contrary, plasticity driven by slip would actually be much more difficult to reliably achieve because of the larger and more complex local displacements. Phenomena like diffusion, where there is significant atomic displacement, may require very large number of replicas to capture intermediate stages

**Table 1:** Slip vector magnitudes of atoms during various stages of twin nucleation process. These atoms are located at different regions within the simulation box. 12 neighboring atoms are considered for each atom in this calculation.

| Type of atom | Atom ID | Slip vector magnitude in Angstorms | | | | |
|---|---|---|---|---|---|---|
| | | $5^{th} - 6^{th}$ | $11^{th} - 12^{th}$ | $17^{th} - 18^{th}$ | $23^{rd} - 24^{th}$ | Average |
| Untwinned | 42847 | 0.0033 | 0.0014 | 0.0047 | 0.0036 | 0.00325 |
| | 5075 | 0.0328 | 0.028 | 0.0064 | 0.0082 | 0.01885 |
| | 48092 | 0.0016 | 0.0016 | 0.0016 | 0.0016 | 0.0016 |
| | 31713 | 0.0034 | 0.0034 | 0.0033 | 0.0034 | 0.00338 |
| Twinned | 19303 | 0.5502 | 0.0042 | 0.0103 | 0.0097 | 0.1436 |
| | 5005 | 0.0304 | 0.7076 | 0.0538 | 0.0084 | 0.20005 |
| | 29250 | 0.0126 | 0.0828 | 0.013 | 0.0122 | 0.03015 |
| | 22588 | 0.2046 | 0.013 | 0.0091 | 0.0086 | 0.05882 |
| GB | 14613 | 0.235 | 0.8761 | 0.6396 | 0.0153 | 0.4415 |
| | 429 | 0.0303 | 0.9293 | 0.1633 | 0.0141 | 0.28425 |
| | 40634 | 0.061 | 0.4535 | 0.1648 | 0.0202 | 0.17488 |
| | 32213 | 0.1848 | 2.8045 | 0.48871 | 0.0557 | 0.91093 |

Some values of the energy barrier and energy relaxed obtained by us are unreasonable since some of them are negative and some are large. However, we expect that the relative comparisons between different grain boundaries should still hold and be useful. This distinction between quantitative and qualitative reliability is often seen in MD simulations. Many results are published with a yield over 2 GPa although both Mg and Ti yield at a much lower value. This discrepancy is usually attributed without much thought to the drastic difference in scales between MD and experimental data. However, inaccuracy in the potentials and the lack of proper pre-exisiting defects in simulations also play a role which is not fully quantified. This inaccuracy has not detracted from the value in such simulations of observed mechanisms and relative effects of boundary and loading conditions. Our data is probably not quantitatively reliable to compare to experiments, but we expect that it is fully capable of good comparative conclusions and extrapolating, for example, what textures of a polycrystal will develop more twin nuclei.

If we look at Figure 4, and 5, then we can see that in the middle section of the graph, increase in energy of replicas is more than the work done. This just means that the replicas are absorbing energy from the surrounding in the middle of the transition which is released at the end in the form of plastic energy. The atoms have certain amount of kinetic energy with them which can be utilized at the beginning to get twin nucleation. This argument can be used to say that low temperature can suppress twinning. Twinning is usually described as athermal, but this is more a description of propagation than nucleation, in which temperature clearly must play a role.

**Conclusion and Further Works**
We determined the required stress for stabilizing a twin nucleus on a symmetric tilt grain boundary is proportional to the GB energy. Mathematically,

$$\sigma_{stabilizing} = k * GB_{energy} \quad\quad\quad 5)$$
$$\approx \frac{10}{3} * GB_{energy}$$

where, $\sigma_{stabilizing}$ has units of MPa, $GB_{energy}$ has units of $mJm^{-2}$ and the constant k has units of $MNmJ^{-1}$. This equation gives a simple relation between GB energy and the stress that was required to stabilize the corresponding twin nucleus. Twin nucleation requires higher stress than twin growth and propagation. We hypothesize that the nucleating stress is some constant times the stabilizing stress. This means, the CRSS for twin nucleation is ultimately a multiple of GB energy. Mathematically,

$$CRSS = c * \sigma_{stabilizing} \quad\quad\quad 6)$$

where, c is any unitless constant.

From our both qualitative and quantitative results, we can draw some important conclusions:
- Higher stress can provide better stabilization of twin embryo.
- In case of STGBs, misorientation angle in the range of 30-60° tilt seems to be the most favorable for twin nucleation.
- Our results suggest that there is a linear correlation between GB energy and stress required to nucleate twin embryos.
- The most stable {10-12} twin embryo we observed formed on the {10-11} twin boundary. This supports the prevalence of double twinning in experimental data.

Twinning is a complex phenomenon and NEB helps overcome this complexity. Because it is a relaxation technique using both initial and final states, there is no timescale dependence. This makes it a really powerful technique. NEB can also be implemented in conjunction with Kinetic Monte Carlo (KMC) to study phenomena like diffusion. KMC requires reaction rate (k) which can be obtained from Arrhenius equation once we have the information on energy barrier. Thus, NEB can act as a precursor to KMC.

**Acknowledgement**